\documentstyle[12pt,aaspp4]{article}
     \raggedright

     \newcommand{\Msun}{\thinspace\hbox{$\hbox{M}_{\odot}$}}
     
     \newcommand{\Lsun}{\thinspace\hbox{$\hbox{L}_{\odot}$}}

\received{}

\begin{document}

     \title{\it INTEGRAL Spectroscopy of IRAS 17208-0014: Implications
     for the Evolutionary Scenarios of Ultraluminous Infrared
     Galaxies\altaffilmark{1}}


     \author{Santiago Arribas\altaffilmark{2,3} \& Luis
     Colina\altaffilmark{4}}



\altaffiltext{1}{Based on observations with the William Herschel
Telescope operated on the island of La Palma by the ING in the Spanish
Observatorio del Roque de los Muchachos of the Instituto de
Astrof\'{\i}sica de Canarias.  Based also on observations with the
NASA/ESA Hubble Space Telescope, obtained at the Space Telescope
Science Institute, which is operated by the Association of Universities
for Research in Astronomy, Inc. under NASA contract number NAS5-26555.}

\altaffiltext{2}{Space Telescope Science Institute, 3700 San Martin Drive, Baltimore, USA (arribas@stsci.edu). Affiliated with the Space Telescope Division, Research and Science Support Department of ESA.}
\altaffiltext{3}{On leave from the Instituto de Astrof\'\i sica de Canarias --
Consejo Superior de Investigaciones Cient\'{\i}ficas (CSIC)}
\altaffiltext{4}{Instituto de Estructura de la Materia, Consejo Superior de Investigaciones Cient\'{\i}ficas (CSIC), Serrano 121, 28006 Madrid, Spain (colina@isis.iem.csic.es)}


     \begin{abstract}

New integral field optical fiber spectroscopy obtained with the
INTEGRAL system, together with archival {\it {\it HST}} WFPC2 and
NICMOS images, have been used to investigate the ultraluminous infrared
galaxy IRAS 17208$-$0014, one of the coldest and most luminous objects in the
IRAS 1 Jy sample.

We have found that the optical nucleus is not coincident with the
true (near-IR and dynamical) nucleus, but it is displaced by 1.3 kpc
(1.5$''$) from it. As a consequence, the previous optical spectral
classifications for the nucleus of this galaxy have to be revised and
changed from HII to LINER. The ionized gas emission is concentrated
around the optical nucleus, where a young (5-6 Myr), massive (3$\pm$1
$\times$ 10$^8$ $\Msun$), and luminous (6$\pm$2 $\times$ 10$^{10}$$\Lsun$)
starburst is detected. Contrary to what is found in {\it dynamically
young} ULIRGs, no strong line emission tracing star-forming regions, or
tidal-dwarf galaxies, is detected in the inner parts of the tidal
tails.

The 2D gas velocity field identifies the {\it optically faint} K-band
nucleus as the dynamical nucleus of the galaxy, and shows
that the 3 kpc, tilted (i $\sim$ 35 degree) disk is rotating at
$\Delta$Vsin$i$= 250 km s$^{-1}$. Radial motions of gas are found along
the minor kinematic axis which, according to the geometry of the
system, are well interpreted as inflow perpendicular to the inner
disk.  The existence of such inflows supports the idea that, as a
consequence of the merging process, gas is channeling from the
external regions, several kpc away, into the nuclear regions where the
massive starburst reported above is taking place.

The kinematical, morphological, and photometric evidence presented here
supports the idea that in IRAS 17208-0014 we are witnessing a luminous,
cool ULIRG which is at the final coalescence phase of a system composed
of two spirals with m $\leq$ m$^*$, a mass ratio of $\sim$ 2:1, each
consisting of a disk+bulge internal structure, that have been involved
in a prograde encounter. This system will most likely evolve into an
intermediate-mass ($\sim$ L$^*$) elliptical. The multifrequency
empirical evidence gathered so far shows no trace of a luminous QSO,
and indicates that starbursts dominate the energy output in this
galaxy. Therefore IRAS 17208$-$0014 does not follow the behavior
expected in the 'ULIRG to QSO' evolutionary scenario proposed by
Sanders et al., but support the one recently
proposed by Colina et al, where two low mass disk galaxies would
produce luminous cool ULIRGs that would not evolve into a QSO phase.

The present study illustrates some caveats to bear in mind when studying
high-z galaxies lacking 2D spectral information of adequate linear
resolution, and shows that near and mid-IR integral field spectroscopy
is needed to derive the relevant astrophysical quantities.

     \end{abstract} 

\section {INTRODUCTION}

Ultraluminous Infrared Galaxies (ULIRGs) are the most luminous objects
in the local universe, with bolometric luminosities L$_{bol} \approx$
L$_{IR} \geq 10^{12}$ L$_{\odot}$. Having large amounts of gas and
dust, they are undergoing an intense starburst activity triggered by
interactions and/or mergers. This starburst activity is believed to be
their major energy reservoir (Genzel et al. 1998), though the relative
importance of the AGN phenomenon as energy power in ULIRGs is under
debate (e.g. Sanders and Mirabel 1996). Sanders et al. 1988 suggested
an evolutionary sequence according to which ULIRGs are the precursors
of optical quasars. In this scenario, {\it warm} ULIRGs (i.e.
$f_{25_\mu}/f_{60_\mu}>$ 0.2) are in a later evolutionary stage than
{\it cool} ULIRGs ($f_{25_\mu}/f_{60_\mu}<$ 0.2). High spatial
resolution imaging and nuclear spectroscopy have found evidence in
support of this scenario (e.g. Surace et al. 1998; Veilleux, Kim,
Sanders 1999 and references there in). On the other hand, Genzel et et
al (2001) have found that most ULIRGs seem to be evolving towards
intermediate-mass (L*) elliptical galaxies, suggesting that ULIRGs
mergers are ellipticals in formation. However, details on how the
ULIRGs evolve remain very uncertain despite the  progress made
recently in characterizing merger evolution on the basis of
theoretical models (see Mihos \& Hernquist, 1996; Naab \& Burkert, 2001
and references therein). These studies show that gas dynamics and star
formation in mergers are closely related.  They also indicate that the
internal structure of the merging galaxies as well as their mass ratio
are key factors in determining the characteristics of ultraluminous
infrared galaxies.

>From an observational point of view, detailed studies of the complex
kinematics and ionization structures of these objects demand the
analysis of their spectral properties in two dimensions (see also Murphy et al. 2001). However, only
a very limited number of this type of studies exists so far (e.g. Mihos
$\&$ Bothun 1998; Wilman, Crawford, Abraham 1999).  We have recently
started a program aimed at studying a representative sample of ULIRGs
combining Integral Field Spectroscopy (IFS) and high resolution {\it
HST} imaging. Some of the main results obtained so far for other ULIRGs
such as Mrk 273, IRAS 12112+0305, IRAS 08572+3915, Arp 220, and IRAS
15206+3342 can be found elsewhere (see Arribas \& Colina 2002, and
references there in).

In this paper, we present new INTEGRAL spectroscopy and archival {\it
HST} imaging of IRAS 17208-0014. This is an
ultraluminous infrared galaxy with some interesting properties. Despite
being the second most luminous galaxy in the sample studied by Kim et
al. (1995) (L$_{bol} = 10^{12.4}$ L$_{\odot}$), it does not show
evidence of an AGN. In particular, i) it has been reported as a case of a
HII/starburst excitation in the optical (Veilleux et al. 1995), and in
the mid-infrared (Lutz,  Veilleux \& Genzel 1999),  ii) the nucleus in
the near and mid infrared images is extended ($\sim$ 2''; Scoville et
al.  2000; Soifer et al.  2000), and iii) the near infrared search for
a hidden broad-line indicated no signs of BLR (Goldader et al. 1995).
{\it HST}-NICMOS observations  (Scoville et al. 2000) have revealed the
presence of at least 18 clusters of star formation, mainly concentrated
in the inner regions, where extinction is high. The presence of a
rotating disk of molecular gas has been inferred from CO observations
(Downes \& Solomon 1998). The nucleus also has one of the strongest
known OH megamasers, and a radio continuum source of $\sim$ 250 pc in
size (Martin et al.  1989).

Ground based optical imaging of IRAS 17208-0014 (Murphy et al. 1996;
Duc, Mirabel, Maza, 1997) has revealed a disturbed and irregular system
with two prominent tidal tails towards the SE and NW extending to
distances larger than 20 kpc. This morphology, together with the fact
that near infrared images show a luminosity profile typical of
ellipticals, have lead to the suggestion that this is a case where an
elliptical galaxy is being formed after a disk-disk collision (Duc,
Mirabel, Maza, 1997). Furthermore, IRAS 17208-0014 is located near the
Fundamental Plane (FP), in a region populated by fast rotating
ellipticals and lenticulars with intermediate mass, and disky isophotes
(Genzel et al 2001).  This result also suggests that the merger
is in a relatively advanced phase, and will eventually lead to an
elliptical of the above mentioned characteristics.

The new optical INTEGRAL spectroscopy and {\it HST} imaging presented
here (Sect. 2) will be used to characterize the stellar structure
(Sect. 3.1), the ionized gas morphology and excitation (Sect. 3.2), the
properties of the extra-nuclear starburst (Sect. 3.3) and the 2D
kinematics of the ionized gas (Sect. 3.4) in this galaxy. On the basis
of these results and previous published studies, we will discuss in
Section 4 the properties and evolutionary phase of the merging system,
as well as the implications for studies of high-z galaxies. Finally, in
Section 5, the main results are summarized.

At the assumed distance of 183 Mpc for IRAS 17202-0014, one arcsec
corresponds to a linear size of 888 pc. Throughout the paper, a Hubble
constant of 70 kms$^{-1}$ Mpc$^{-1}$ is assumed.

\section{OBSERVATIONS AND DATA REDUCTION}

Integral field spectroscopy of IRAS 17202-0014 was obtained with the
INTEGRAL system (Arribas et al. 1998) connected to WYFFOS (Wide Field
Fiber Optic Spectrograph; Bingham et al. 1994) on the 4.2m William
Herschel Telescope during 1999 April 3.  The bundle of fibers SB2
(Standard Bundle number 2), consisting of 219 fibers, each 0.9$^{''}$
in diameter, was used. This is arranged such that 189 fibers cover a
rectangular area of 16.5$^{''}$ $\times$ 12.3$^{''}$ (14.6 kpc $\times$
10.9 kpc at the distance of IRAS 17202-0014) while 30 additional
fibers, forming a ring of 90$''$ in diameter concentric with the
rectangle, measure simultaneously the sky background (see Arribas \&
Mediavilla 2000 for further details on integral field fiber
spectroscopy). The spectra were taken using a 600 line/mm grating with
an effective resolution of 4.8 \AA\ and covering the 5000-7900
\AA\ spectral range. The total integration time was 7800 seconds split into
4 separate exposures, with seeing of about 1.1$^{''}$.

Data reduction followed the standard procedures applied to spectra
obtained with two-dimensional fiber spectrographs (see, for instance,
Arribas \& Mediavilla 2000, and references therein).  As an example, the
two-dimensional distribution of the reduced spectra in the range of the
redshifted H$\alpha$+[NII]+[SII] lines (i.e. 6792 - 7045 \AA) are
presented in Figure 1. H$\beta$ and the [OIII]$\lambda\lambda$4959,5007
appeared conspicuously in only four spectra: 102, 103, 104, and 107
(not shown).

For each spectrum, the radial velocities and velocity dispersions were
measured by adjusting Gaussian functions to the observed emission line
profiles using the DIPSO package (Howarth $\&$ Murray 1988).  The
uncertainties associated with the individual velocities are estimated
to be in the order of $\pm$ 15 km s$^{-1}$.  The velocity dispersion
values presented in this paper are corrected from instrumental profile
and redshift.

The archival {\it HST} images of IRAS 17208-0014 obtained in the I-band
(F814W filter, WFPC2) and K-band (F222M, NIC2) were originally taken as
part of programs 6346  (PI: K. Borne) and 7116 (PI: N. Scoville),
respectively. The   total exposure time for the F814W image was 800
seconds divided into two identical  exposures of 400 seconds each,
using the CR-SPLIT option. The data were recalibrated  at the time of
dearchiving using on-the-fly calibration software and the best
reference calibration files available. The NICMOS F222M image, first
analysed and published in Scoville et al. 2000, with a total
integration time of 288 seconds, was recalibrated in October 2002 with
{\it calnica} and {\it calnicb} with the best available reference files.

\section{RESULTS}

\subsection {Stellar structure and differential extinction effects}

The INTEGRAL red continuum  image taken with a narrow 7029-7068
\AA\ window (Fig. 2, bottom-left panel) has a good morphological
resemblance to the WFPC2 I-band image (Fig. 2, upper-left panel).
This allows us to establish the absolute positioning of the INTEGRAL
maps with an uncertainty of about 0$\farcs3$ with respect to the {\it
HST} reference frame.  In addition, the WFPC2 and NICMOS images could
be placed on the same coordinate system using point-like structures
present in both images.

The morphology of the outer low-surface brightness region in the red continuum
images is like that of a perturbed elliptical. However, inwards, the
presence of spiral arm-like structures are rather  conspicuous.  The
innermost star-forming clusters previously identified in the near-IR
with NICMOS (Scoville et al., 2000; see also Fig. 3) are also detected
in the WFPC2 I-band image. Leaving aside the contribution of the
circumnuclear clusters, the inner isophotes in the infrared images
define an elliptical structure, suggesting the presence of a tilted
disk with i= 35$\pm$15 degrees and 3$\pm$0.5 kpc in diameter (Fig. 3).

There is a general consensus that ULIRGs are galaxies where most of
their active regions, either massive starbursts or AGN, are enshrouded
in dust. However, the dust distribution, although concentrated towards
the nuclear regions, i.e.  inner few kpcs, is very patchy and therefore
differential extinction effects play a major role in interpreting the
results (e.g. Colina et al. 2001). In IRAS 17208$-$0014, the most
relevant effect of the differential extinction is that the true nucleus
of the galaxy does not coincide with the region identified as the
nucleus in the optical, but it is located about 1.5 arcsec (1.3 kpc)
southwest of it (compare Fig.2 upper-left with Fig. 3). In the
direction of the optical nucleus, the hydrogen recombination lines (see
next section) in the INTEGRAL spectra indicate an extinction A$_{v}$ of
about 5.5 magnitudes, in good agreement with previous measurements (see
Table 1).  No H$\beta$ is detected towards the near-IR nucleus.
However, assuming that extinction is in a foreground screen (i.e. not
mixed with the stars), and using the law derived by Rieke and Lebofsky
(1985) an extinction of up to A$_{v}$ = 8 magnitudes is inferred from
the infrared colors (Scoville et al 2000).  However, this could be a
lower limit. Scoville et al.  pointed out that the nuclear power source
in IRAS 17208$-$0014 could be buried even at near-infrared
wavelengths.

A secondary emission peak in the optical, located at about 3$''$ (2.7
kpc) southeast of the optical nucleus, is rather clear in both the
INTEGRAL and the WFPC2 images. This region could, in principle, be
identified as the second nucleus of a merging pair. However, its flux
loses strength at longer infrared wavelengths, and nearly disappears
in the 2.2 $\mu$m map (Fig. 3), suggesting that its optical emission is
associated with a relatively young population located on one of the
spiral arm-like structures.

 \subsection{Ionized gas morphology and excitation structure}

The structure of the warm ionized gas is traced here by the H$\alpha$
map (Fig. 2, central upper panel). This image shows a bright nucleus
and a low intensity extended emission of at least 8 kpc elongated in
the SE - NW direction. The structure of the outer envelope is
reminiscent of that of the stellar  component, though it does not have a
one-to-one correspondence.  In fact, the ionized gas emission is more
compact and it extends towards the North, where a tidal tail is
observed in deep optical continuum images (see, Murphy et al. 1996;
Duc, Mirabel, Maza, 1997).  However, no strong line emission  tracing
star-forming regions or tidal-dwarf galaxies is observed in the inner
parts of the tidal tails, as found in some double-nuclei  ULIRGs (e.g.
IRAS 12112+0305:  Colina et al. 2000; IRAS 08572+3915:  Arribas,
Colina, Borne 2000; IRAS 14348$-$1447: Monreal et al., in preparation).

The ionized gas in the optical nucleus exhibits excitation conditions
characteristic of HII regions (see Table 1 and Fig. 2). Strong
H$\alpha$, relative to [NII], is also detected towards the south of the
optical nucleus.  However, apart from these regions, the [NII] emission
is at least as intense as H$\alpha$ in the rest of the regions (see
Fig. 1).  Further support for this dichotomy in the excitation
conditions comes from the ratio [SII]/H$\alpha$, which has a minimum in
the optical nucleus, and is considerably  higher elsewhere (see Fig. 1
and compare, for instance, spectra 103 and 100). H$\beta$ and [OIII]
are detected in only a few spectra outside the optical nucleus, with a ratio of [OIII]/H$\beta$ $\sim$ 0.5. Note that for a [OIII]/H$\beta$ ratio lower
than 1,  the values of the [OI]-, [NII]-, and [SII]-to-H$\alpha$ ratios,
almost unaffected by  internal extinction, are good discriminators
between LINER and HII ionization (Veilleux $\&$ Osterbrock, 1987).
Therefore, our new INTEGRAL spectroscopy shows that outside the optical
nucleus, the excitation properties of the ionized gas are
characteristic of LINERs (see Table 1 and Fig. 2). This includes the
near-IR nucleus (i.e. true nucleus, see $\S$ 3.4), which is
therefore classified as a LINER (Table 1). Since previous
classifications were based on long-slit spectroscopy of the optical
nucleus (Veilleux et al, 1995), the classification of the
nucleus of IRAS 17208$-$0014 has to be revised and changed to LINER.
This new classification is not in contradiction with the classification
as a starburst based on ISO spectroscopy (Lutz, Veilleux, and Genzel,
1999), especially if we take into account the different spatial
resolution.  On the contrary, as pointed out by these authors, rather than indicating the presence of an active nucleus, the LINER spectra in ULIRGs are more likely due to shocks produced by compact and massive
starbursts.

\subsection{Massive extranuclear starburst}

The properties of the optical emission peak, i.e. the optical nucleus,
indicate that a  massive, young, starburst is taking place
in an extranuclear region (see $\S$ 3.4). The strong hydrogen recombination lines
with equivalent widths of 94 $\pm$ 10\AA\  and 13 $\pm$ 3\AA\  for
H$\alpha$ and H$\beta$, respectively, indicate the presence of a young
burst of about 5 to 6 million years.  Considering an age of 6 Myrs, the
dereddened H$\alpha$ luminosity (see Table 1), and the WFPC2 I-band
absolute magnitude (M$_I$= $-$21.7, if A$_V$ = 5.5 mag. is assumed) can
be used to obtain two independent estimates of the total mass in the
burst. Assuming the predicted H$\alpha$ and I-band luminosity for an
instantaneous burst characterized by a Salpeter initial mass function
and stellar masses in the 1 to 100 $\Msun$ range (STARBURST99;
Leitherer et al. 1999), the two estimates agree and give a total mass
of 3 $\pm$1 $\times$ 10$^8$ $\Msun$. The bolometric luminosity of such
a young and massive starburst is 6 $\pm$2 $\times$ 10$^{10}$ $\Lsun$.  Thus,
although very massive, this starburst still produces a small fraction
of the energy output of IRAS 17208$-$0014, equivalent to 2.5\% of its
IR luminosity.  The region where this starburst is located has a size
of about 0.9 kpc in diameter. Therefore, it is likely that this
starburst, as suggested by the {\it HST} images, does not form a single
entity but is distributed in several less massive clusters. In this
respect, most of the bright clusters identified in the near-IR are
located in this region (Scoville et al. 2000).  In particular, the two
first-ranked K-band clusters alone (clusters 1 and 9 in Scoville et
al.  2000) produce a combined bolometric luminosity of 2 $\times$
10$^{10}$ $\Lsun$, i.e. a factor of three less than the luminosity of the
starburst, if the same age and extinction as measured for the entire
region is assumed.

\subsection{Kinematic tracing of the dust-enshrouded dynamical nucleus
and inner gas disk}

The velocity field of the ionized gas inferred from the H$\alpha$+[NII]
lines is shown in Figure 2 (bottom-right panel). The largest velocity
gradient is not found across the optical nucleus, but across the
infrared nucleus (see also Fig. 3, bottom). This indicates that the
dynamical nucleus is in positional agreement with the infrared
nucleus.  Therefore, even if the nuclear power source is buried at near
infrared wavelengths, the infrared nucleus indicates its direction.
Note that the highly symmetric pattern of the ionized gas velocity
field allows us to accurately locate the position of the true
nucleus using optical spectroscopy (i.e without the need of near-IR
observations).

The peak-to-peak velocity difference corresponds to $\Delta$Vsin$i$=
250 km s$^{-1}$ along PA $\sim$ 130 degrees.  This direction is in good
agreement with that of the major photometric axis inferred from the
inner isolines in the NICMOS images (excluding the effects from the
clusters; see Fig. 3). The agreement between the photometric and
kinematic axes suggests that we are observing a rotating disk of
ionized gas. The kinematic axes of this disk are also in good
positional agreement with those of the molecular (CO) disk found by
Downes and Solomon (1998). It is also worth noting that the velocity
field does not show any evidence for the presence of a second nucleus.

Outside the inner disk the iso-velocities twist, having a characteristic
S-shape.  Although this may indicate warps (e.g. Mediavilla et al.
1991), we believe it is due to the transition from inner-disk to
outer-envelope seen in projection, as suggested by fact that the
isovelocity distortions appear at the edge of the inner disk.

The velocity dispersion map indicates relatively large velocity
dispersions over most of the observed region (i.e. FWHM $\sim$ 300
km/s).  Within uncertainties, the infrared nucleus is in positional
agreement with a local maximum ($\sim$ 450 km/s) in the velocity
dispersion map (Figure 3), which in turn is well centered with the
kinematic center of rotation. Therefore, there is little doubt that the
dynamical nucleus of the galaxy is located there.  Another local
maximum in the velocity dispersion map is observed towards the SW at
about 3 arcsec (2.7 kpc). The mean velocity of this region is
redshifted by about 200 km/s with respect to the systemic, as can be
observed in the velocity field map. These redshifted velocities are
found in a position near the kinematic axis, where mainly systemic velocities
are expected for a purely rotating disk. This indicates the presence of
a kinematically distinct component.

\section {DISCUSION}

\subsection {Tidal-induced inflows}

In principle, the kinematically distinct component reported in \S 3.4
could indicate the presence of large-scale outflows originated by
AGN-driven jets (Colbert et al. 1998 and references therein). However,
the fact that no evidence for either an AGN or a jet has been
reported for this object makes this explanation rather speculative.
Superwinds (Heckman et al. 1990) could also be, in principle, a
possible explanation.  However, if the spiral-like arms observed in
optical images are trailing, the INTEGRAL velocity field of the inner disk (i.e. redshifted velocities in the SE; blueshifted velocities in the NW) indicates that the closer part of the disk is in the NE. With this geometry, and
assuming an inclination of 35 degrees (as suggested by the inner
isolines in the NICMOS K image), two outflowing components are expected
to be seen in projection over the SW and another two over the NE (see the
geometry described in Figure 17 of Heckman et al.  1990). In the SW,
the component closest to us should have blueshifted velocities. The
farthest component seen in projection over the SW could have blueshifted
or redshifted velocities depending on the opening of the outflow cone.
The fact that we only observe one extra component (out of four expected)
which also has relatively strong redshifted velocities along the
SW part of the minor axis makes the superwind scenario very
unlikely in the case of IRAS17208-0014.

Alternatively, the kinematically distinct component could be associated
with streaming motions of gas related to the merging process. The fact
that these motions are found along the minor axis suggests that they
are perpendicular to the disk of gas. Taking into account the geometry
of the inner disk (i.e. with NE being the closer part, i=35 degrees),
the observed redshifted velocities along the SW part of the rotation
axis imply inflow.  (Conversely, outflows along the SW part of the
rotation axis would imply approaching velocities, i.e. a blueshifted
component). Note that the rotation axis is a canal where mass can
travel efficiently towards the innermost parts of the galaxy without
the limitations imposed by the conservation of angular momentum. In
short, the extra redshifted component found along the SW part of the
minor axis can be explained as the result of an inflow motion
perpendicular to the disk. Even if the inflows may lose efficiency at
the innermost regions (i.e., the inner kpc), their existence supports
the idea that, as a consequence of the merging process, gas is
channeling from the external regions, several kpc away, into the
nuclear regions where the massive starburst described in $\S$ 3.3 is
taking place. The presence of strong inflows and associated star
formation are expected during the final coalescence phase of two disk
galaxies with bulges (Mihos and Hernquist, 1996). IRAS 15206+3342
(Arribas \& Colina 2002) is another case where these type of motions
are clearly observed.

\subsection{IRAS 17208$-$0014 Near-IR Luminosity: Extinction
Corrections for ULIRGs}

The measurements of the luminosity of ULIRGs in the near-infrared and
the internal extinction corrections to be applied are not well
defined.  Some authors (Genzel et al. 2001) argue that extinction over
the entire size of these galaxies is very large and therefore
corrections to the integrated apparent magnitudes, even in the K-band,
are not negligible, i.e. mean corrections of 0.7 mag have to be
applied. Others (Colina et al. 2001) indicate that dust distribution is
very patchy in ULIRGs and therefore while internal extinction towards
the nucleus can be high, the mean extinction over the entire ULIRG is
much less due to large differential extinction gradients.  In the
following, we discuss the differential extinction in IRAS 17208$-$0014
and show how this extinction should be taken into account in order to
obtain a reliable near-IR luminosity of this galaxy, and of ULIRGs in
general.

In IRAS 17208$-$0014, NICMOS colors (Scoville et al. 2000) and the
INTEGRAL spectroscopy presented here (see $\S$3.1) indicate visual
extinctions of about 8 and 5.5 mag towards the near-IR
nucleus and the extranuclear starburst (i.e. ``optical nucleus''),
respectively. However, the mean extinction across the entire galaxy
appears to be much less as indicated by the bluer near-IR colors
(Scoville et al. 2000) when increasing the size of the aperture from
1$\farcs$1 to a 11$\farcs$4 diameter (i.e. from 1 to 10 kpc at the
assumed distance for IRAS 17208$-$0014).  Stellar populations cover
a limited and well defined range of near-IR colors with  almost
constant J$-$H of about $+$0.8 and H$-$K of $+$0.35 for starbursts
older than 10 Myr, and slightly bluer colors for younger starbursts
(STARBURST99; Leitherer et al. 1999).  Therefore, the observed,
integrated colors can be used to estimate the mean extinction within
the galaxy. For IRAS 17208$-$0014, the NICMOS J$-$H and H$-$K colors
measured within an aperture of 11$\farcs4$ or 10 kpc (i.e. covering the
field-of-view of our INTEGRAL spectra) are $+$1.3 and $+$0.6,
respectively, and therefore the mean extinction (A$_V$) is about 2.5
mags, much less than for the nuclei. Therefore, the extreme extinctions
measured in the nuclear regions ($\sim$ 1 kpc) through small apertures
can not be applied to the entire galaxy.

This fact is obviously important when computing the absolute near-IR
magnitudes of ULIRGs in general, and of this galaxy in particular. The
observed apparent H-band magnitude of 12.09 in a region of 10 kpc in
diameter, i.e. including the nucleus and host (Scoville et al. 2000),
corresponds to an absolute magnitude of M$_H$= $-$24.22 (i.e. an L$^*$
galaxy) if no reddening correction is applied.  If the mean extinction
of A$_V$= 2.5 mag is considered for the entire galaxy,  the
corresponding luminosity of the system is 1.5L$^*$.

\subsection{Characteristics of the Merger}

The morphological and kinematical properties of IRAS 17208$-$0014
indicate that this system is in an advanced phase  of the merging
process between two gas-rich galaxies. The predicted strong inflow (and
associated star formation) in this phase is well detected in the case
of IRAS 17208$-$0014, as we have discussed in $\S$ 4.1 (and $\S$ 3.3).
High-resolution NICMOS images (Scoville et al. 2000, see also Fig. 3)
and INTEGRAL two-dimensional kinematics do not show evidence for a
second nucleus. The massive young clusters are preferentially located
in the circumnuclear regions (i.e. inner 1-2 kpc) with no signs of
star forming regions along (the observed inner parts of) the tidal
tails, like those found in dynamically younger systems where two
distinct nuclei are still present (e.g.  IRAS 12112+0305, Colina et al,
2000; IRAS 08572+3915, Arribas, Colina, Borne 2000; IRAS 14348$-$1447,
Monreal et al. in preparation).  Further morphological characteristics,
e.g.  $\sim$ r$^{1/4}$ intensity profile in the near-IR (Scoville et
al.  2000),  also indicate that IRAS 17208$-$0014 is indeed an advanced
merger/elliptical in formation.

The fairly regular disk kinematics shown in our velocity field (as well
as the good alignment of the kinematic and photometric axes) points to
an unequal-mass merger according to the numerical simulations by Bendo
and Barnes (2000) and Naab and Burkert (2001). This is supported by
recent measurements of the central V$_{rot}$/$\sigma$ ratio from
stellar kinematic data (Genzel et al. 2001). The value
V$_{rot}$/$\sigma$ of 0.48 indicates that the merger product is a
relatively fast rotator and therefore that the mass ratio of the two
progenitor galaxies involved in the merger should be $\sim$ 2:1.
Moreover, the relatively long (though faint) tidal tails in
IRAS 17208-0014 suggest that the encounter could have been prograde
(Dubinski, Mihos \& Hernquist 1999).

The estimated dereddened H-band luminosity of IRAS 17208$-$0014
corresponds to 1.5L$*$ (see $\S$4.2). This result by itself does not
imply that the system is the merger of two massive spirals  (i.e.
$\geq$ m$^*$). In fact, young (6 to 10 Myr) starbursts have H-band M/L
ratios a factor of 50 lower than that of old 1 Gyr stellar populations. In
other words, the same H-band luminosity will be produced by an old
bulge population and by a 50 times less massive young starburst.  As an
example, the massive extranuclear starburst of 3 $\times$ 10$^8$
$\Msun$ (see $\S$3.3) alone produces an H-band luminosity of 0.1
L$^*$.  Moreover, a continuous 100 $\Msun$ yr$^{-1}$ star formation
over 10-20 million years, inferred to be happening in ULIRGs in
general, produces an L$^*$ luminosity in the H-band. Recent stellar
kinematical measurements support low M/L values. In fact, IRAS
17208$-$0014 is located near the region of the Fundamental Plane (FP)
populated by intermediate-mass ($\sim$ L$^*$) ellipticals, and its
dynamical mass corresponds to 1.1 $\times 10^{11} \Msun$, or 0.8$m^*$,
from stellar velocity dispersion measurements (Tacconi et  al. 2002).
Although the gas kinematics may not necessarily trace the dynamical mass,
we note that the present gas velocity dispersions are only somewhat
lower than the stellar values which also suggests a mass lower than m*
for the system. These results therefore support the scenario proposed
by Colina et al.  (2001) where cool ULIRGs like  IRAS 17208$-$0014
would be produced during the coalescing of two low-mass disk galaxies.

In summary, the kinematical, morphological, and photometric evidence
presented here supports the idea that we are witnessing in IRAS
17208-0014 a luminous, cool ULIRG which is at the final coalescence
phase of a system composed of two low mass spirals (i.e. $\leq$ m$^*$),
with a mass ratio of 2:1, each consisting of a disk+bulge internal
structure, that have been involved in a prograde encounter. This system
will most likely not evolve into a luminous QSO (see next section), but
into a fast rotating elliptical ($\sim$ L$^*$).

\subsection{IRAS 17208$-$0014: A Cool ULIRG not Evolving into a
Luminous QSO}

According to the evolutionary scenario first proposed by Sanders et al.
(1988), cool ULIRGs would evolve into warm ULIRGs, and finally into
QSOs. This scenario implies that cool ULIRGs should be found
preferentially among galaxies in the initial and intermediate phases of
the merging process while the galaxies involved in the collision still
preserve much of their individual identities, or are separated enough
that their nuclei have not merged yet. On the other hand, warm ULIRGs
should be found in late phases and therefore preferentially  among
close pairs, or even single-nucleus galaxies, while QSOs should be
located  exclusively in compact, single nucleus systems.

According to this scenario, IRAS 17208$-$0014, which is at an advanced
 merging phase (see $\S$ 4.3) and has an infrared luminosity  of
 L$_{IR}= 10^{12.4}$ $\Lsun$  (i.e. one of the most luminous ULIRG in
 the 1 Jy sample), should be showing the characteristics of a warm,
 QSO-like ULIRG.  However, quite the opposite, IRAS 17208$-$0014 shows
 no signs for the presence of a QSO: (1) with a $f_{25}/f_{60}$ ratio
 of 0.05, it is one of the coldest  ULIRGs, (2) its near and mid-IR
 emission show the same extended structure of about  2$''$ (Scoville et
 al.  2000; Soifer et al. 2000) with no evidence of a  luminous,
 point-like, nuclear source, (3) there is no trace of a (dominant) AGN
 component from either the optical (this work), or the near (Goldader
 et al.  1995) and mid-IR (Lutz et al.  1999) spectroscopy, and not
 even from the hard X-ray (Risaliti et al. 2000). Moreover, the low
 f25/f60 and F$_{HX}$/F$_{IR}$ (1.25 $\times$ 10$^{-4}$) flux ratios
 are close to those of luminous infrared (i.e. L$_{IR}$ $\le$
 10$^{11}$) starbursts, and unlike the values of luminous infrared
 galaxies with a Seyfert 1 nucleus (about 2 and 0.1, respectively; see
figure 5 in Risaliti et al 2000).  In addition, considering the
dereddened H$\alpha$ luminosity for the entire galaxy (3.4 $\times$
10$^{42}$ erg s$^{-1}$), the ratio
 of hard X-ray to H$\alpha$ luminosity is 0.35, much less than the
 median value of 15 for QSOs, Seyfert 1, and low-luminosity LINER 1
 nuclei (Ho et al.  2001), and similar to the average measured value
 (0.14) in nearby nuclear starbursts (P\'erez-Olea \& Colina 1996).
 Finally, the low radio flux density of 102 mJy at 1.4 GHz (Condon et
 al. 1996) gives an X-ray to radio luminosity ratio of 520, also in
 agreement with the average value of 400 in nearby nuclear starbursts
 (P\'erez-Olea \& Colina 1996).

In addition to these energy arguments, there are also dynamical
arguments indicating that, if IRAS 17208$-$0014 harbors a central
black-hole, it must have a relatively low mass  and therefore not
capable of producing a luminous QSO (M$_{BH}$ $\sim$ 10$^9$
$\Msun$).  The estimated mass of the IRAS 17208$-$0014 black-hole is
2.3 $\times$ 10$^8$ $\Msun$, based on ground-based near-infrared
stellar  velocity dispersion measurements and assuming the correlation
between the black hole masses and bulge velocity dispersion for early
type galaxies also holds for late phase ULIRGs (Tacconi et al. 2002).
The stellar velocity dispersion has been measured over a region  of
about 1 kpc in diameter (0.5 to 1.2 effective radii as mentioned in
Tacconi et al. 2002). The central gas and stellar distribution in
ULIRGs differ dramatically from that of ellipticals and bulges of
spirals due to their large concentrations of high density molecular gas,
and young, massive stellar clusters. Therefore, it is likely that in
ULIRGs, aside from the black hole itself, these stellar and gas
components make a non-negligible fraction of the total mass.  Therefore,
the mass estimate of the nuclear black hole mass should most likely be
considered as an upper limit to the true mass of the black-hole.

In summary, despite its advance dynamical phase, starbursts seem to be the
dominant energy source in IRAS 17208$-$0014 and, if an AGN exists, it
must produce only a minor contribution to the overall energy and ionizing
budget. We note that similar characteristics (i.e. advanced merger
without clear evidence of QSO) have also been found in IRAS 15206+3342
(Arribas and Colina 2002). These cases do not follow the behavior
expected in the 'ULIRG to QSO' evolutionary scenario proposed by
Sanders and collaborators, therefore questioning its universality. On the other
hand, if the alternative scenario proposed by Colina et al.  (2001) is
correct, IRAS 17208$-$0014 represents an example of the merger of two
low mass disk galaxies that will produce a luminous, cool ULIRG but
will never evolve into a QSO phase.

\subsection{Implications for High-z Dust-enshrouded Massive Starburst
Galaxies}

There are three fundamental questions that have to be answered in the
future in order to advance our knowledge  of high-z galaxies: (1) what
is the fraction of AGNs as a function of redshift, (2) what is the mass
distribution of the pre-galactic objects in the early Universe, and (3)
how do pre-galactic objects evolve into present-day galaxies (i.e.
interaction/merging rates, number of colliding objects)?  The results
presented for IRAS 17208$-$0014 illustrate the problems and caveats
that have to be taken into account when investigating high-z dusty
objects and before answering the above mentioned questions.

IRAS 17208$-$0014 is a clear example of a ULIRG where the optical
nucleus, both in continuum and emission lines, does not trace the true
dynamical nucleus of the  galaxy but is displaced by a large amount of
1.3 kpc. Moreover, IRAS 17208$-$0014 shows two strong continuum peaks
in the optical that could naively be associated with  the nuclei of two
galaxies. However, two-dimensional kinematics and K-band imaging show
that none of these optical nuclei are associated with the {\it true
single} nucleus of this galaxy.  Therefore, conclusions about the
morphological properties and structure of the progenitors of high-z
dusty  starbursts, including merging rates, interaction phase, or
number of colliding objects, would depend critically on a reliable
two-dimensional knowledge of the stellar structure and kinematical
properties of the gas.

IRAS 17208$-$0014 is also an example of misleading spectral
classification of its nucleus in the optical.  In the past, due to the
lack of two-dimensional spectroscopy,  this ULIRG has been classified
as an HII, while the true nucleus has a LINER spectrum,  as shown
here.  If this is a common situation, it could have severe
implications. Since the  fraction of AGN in galaxies as a function of
redshift can trace the formation and growth  of massive black-holes at
early epochs, establishing a reliable classification of the  nuclei of
high-z galaxies, in particular of the most common low-luminosity AGN,
is obviously needed.

Regarding its kinematics, the largest velocity gradient and the
velocity dispersion peak in IRAS 17208$-$0014 are not found towards the
optical nucleus, but towards  the optically faint, K-band
nucleus. If the emission peak of the ionized gas is decoupled from the
stellar continuum, or the stellar and gas distribution do not trace the
true nucleus of the galaxy, any mass estimate from optical emission
lines (e.g. H$\alpha$) based on spatially unresolved velocity
dispersion  measurements will be affected by an undetermined error,
depending on the specific object.  IRAS 17208-0014 is not an isolated
case among ULIRGs. Other ULIRGs for which integral field spectroscopy
is available (i.e. Mkr 273:  Colina, Arribas \& Borne 1999; IRAS
12112+0305: Colina et al. 2000; IRAS 08572+3915: Arribas, Colina \&
Borne 2000; Arp 220: Arribas, Colina, Clements 2001; IRAS 15206+3342:
Arribas \& Colina, 2002) do show similar features, most notably the
decoupling between the stellar and ionized gas structure, offsets
between the velocity dispersion peak and the optical light
distribution, and strong differential extinction effects that could
lead to wrong conclusions if investigated without adequate spatially
resolved 2D spectroscopy.

Future near and mid-IR observations of high-z dusty galaxies with large
telescopes will effectively observe the rest-frame optical spectrum
(0.4 - 0.7$\mu$m) for redshifts of three, or larger. If, as
demonstrated in this paper (see also previous papers of this series)
optical IFS data is relevant for adequately interpreting the properties
of local ULIRGs, so near and mid-IR IFS should be for interpreting
high-z galaxies. Although sensitivity and angular resolution (which
worsen with wavelength) may limit these types of studies, a good
understanding of the observational effects from analyzing a
representative sample of local ULIRGs seems crucial in order to grasp
the limitations of poorer linear resolution and 1D studies.

\section{SUMMARY:}

Integral field optical fiber spectroscopy with the INTEGRAL system,
together with archival {\it HST} WFPC2 and NICMOS images, have been
used to investigate the ultraluminous infrared galaxy IRAS
17208$-$0014.  The main results are the following:

1) Differential extinction plays a major role in interpreting the
innermost regions of this system. The optical nucleus (in both the
continuum and line emission) is not coincident with the true
(near-IR and dynamical) nucleus, but is displaced by 1.3 kpc
(1.5$''$) from it. As a consequence, the previous optical spectral
classification of the nucleus of this galaxy has to be revised and
changed from HII to LINER.

2) A mean internal extinction equivalent to 2.5 mag in the visual is
derived for the galaxy. Due to the patchy structure of the dust
distribution, this average extinction is much lower than the values
(5.5 to 8 mag) measured in the nuclear regions.  A dereddened H-band
luminosity of $\sim$ 1.5L* is found for the entire system. Most of this
luminosity is dominated by the young starbursts, not by the old bulge
stellar population of the progenitor galaxies. In fact, in agreement
with recent stellar measurements (Tacconi et al, 2002), our data
suggest that IRAS 17208-0014 is a low mass (i.e. $< m^*$) system.

3) The ionized gas emission is concentrated around the optical nucleus,
where several luminous star-forming clusters are observed in the {\it
HST} images. Our multi-wavelength analysis indicates that the age of the
burst is 5-6 Myr, its mass 3$\pm$1 $\times$ 10$^8$ $\Msun$, and its
bolometric luminosity 6$\pm$2 $\times$ 10$^{10}$$\Lsun$.  This starburst
produces only a small fraction of the energy output of IRAS
17208$-$0014, equivalent to 2.5\% of its  IR luminosity.

4) No strong line emission tracing star-forming regions or tidal-dwarf
galaxies are observed in the inner parts of the tidal tails, in
contrast with those found in dynamically young double-nuclei ULIRGs.
In fact, the secondary nucleus observed in the optical continuum (at
2.7 kpc from the optical nucleus) vanishes at longer wavelengths and
does not have distinct kinematics, excluding the possibility of being
associated with the nucleus of one of the merging galaxies. 

5) The 2D velocity field has identified the dynamical center
of the galaxy as the region coincident with the near-infrared nucleus
detected in the NICMOS images. The velocity field also indicates that
the 3 kpc inner disk detected in the K-image has a regular rotation
pattern ($\Delta$V$sini$= 250 km/s). The kinematic and photometric axes
are well aligned, which suggests that the stellar and gas components
are coupled in this region. No kinematic evidence for a secondary
nucleus is found.

6) Radial motions of gas are found along the minor kinematic axis.
According to the geometry of the system (fixed by the velocity field
and the assumption that the spiral arms-like structures are trailing),
the radial motions are well interpreted as inflow perpendicular to the
inner disk.  The existence of such inflows supports the idea that, as a
consequence of the merging process, gas is channeling from the
external regions, several kpc away, into the nuclear regions where the
massive starburst mentioned above is taking place.

7) The kinematical, morphological and photometric evidence presented
here supports the idea that in IRAS 17208-0014 we are witnessing a
luminous, cool ULIRG which is at the final coalescence phase of a
system composed of two low mass spirals (i.e. $\leq$ m$^*$),  with a
mass ratio of 2:1, each consisting of a disk+bulge internal structure,
that have been involved in a prograde encounter. This system will most
likely evolve into a rotating, intermediate-mass ($\sim$ $L^*$)
elliptical.

8) All the  multifrequency empirical evidence gathered so far indicates
IRAS 17207$-$0014 does not have a luminous QSO nucleus. Starbursts seem
to be the dominant energy source and therefore, if an AGN exists, it
must produce a minor contribution to the overall energy and ionizing
budget. Therefore, IRAS 17208$-$0014 does not follow the behaviour
expected in the 'ULIRG to QSO' evolutionary scenario proposed by
Sanders and collaborators. In fact, the present data (as well as those
of IRAS 15206+3342; Arribas \& Colina, 2002) are in better
agreement with the alternative scenario proposed by Colina et al.
(2001), according to which cool ULIRGs such as IRAS 17208$-$0014 are
the merger of two low mass ($< m^*$) disk galaxies that will never
evolve into a QSO phase.

9) The present study illustrates some caveats to bear in mind when
studying high-z dusty galaxies without 2D spectral information at
adequate linear resolution.  Near and mid-IR integral field
spectroscopy is needed to investigate the nature and evolution of
luminous infrared galaxies in the early Universe.

\acknowledgments

We thank all the staff at the Spanish Observatorio del Roque de los
Muchachos of the Instituto de Astrof\'{\i}sica de Canarias for their
support. We appreciate the careful reading and comments from Ray Lucas,
as welll as those from an anonymous referee.  Luis Colina thanks
STScI for financial support under the Collaborative Visitors Program.
We also thank Luis Cuesta by his help using  GRAFICOS. Financial
support for this work was provided by the Spanish Ministry for Science
and Technology through grants PB98-0340-C01, PB98-0340-C02, and
AYA2002-01055.

\newpage \scriptsize \begin{deluxetable}{ccccccccc} \tablewidth{41pc}
\tablecaption{Ionized gas properties of the optical and near-IR nuclei}
\tablehead{Region& A$_V$\tablenotemark{b}
&F(H$\alpha$)\tablenotemark{c} & EW(H$\alpha$)\tablenotemark{d} &
logL(H$\alpha$)\tablenotemark{e} &
log([\ion{O}{3}]/H$\beta$)\tablenotemark{f} &
log([\ion{O}{1}]/H$\alpha$)\tablenotemark{f} &
log([\ion{N}{2}]/H$\alpha$)\tablenotemark{f} &
log([\ion{S}{2}]/H$\alpha$)\tablenotemark{f} \\ & (mag.) & & (\AA) &
(\Lsun) & & & & \\}

\startdata Opt. Nucleus  &5.5$\pm$0.5& 0.56$\pm$0.1 & 94$\pm$15 & 8.55$\pm$0.1 & $-$0.002$\pm$0.09 & $-$1.60$\pm$0.1 &
$-$0.40$\pm$0.06 & $-$0.51$\pm$0.06 \nl
	      &     &      &    &      & $-$0.10$\pm$0.09 & $-$1.35$\pm$0.1 & $-$0.41$\pm$0.06 &
	      $-$0.56$\pm$0.06 \nl
Opt. Nucleus\tablenotemark{(a)}& 5.4&1.79 & 51 & 9.05 & $-$0.16 &
$-$1.22 & $-$0.35 & $-$0.54\nl Near-IR Nucleus &$\sim$8 & 0.08 & 17 &
8.47 & & $-$0.68 & $-$0.07 & $-$0.12 \nl 
&      &    &      &         &         &         \nl
\tablenotetext{a}{All values taken from Kim, Veilleux, and Sanders (1998) using a slit
width of 2$''$ along the E$-$W direction. According to Kim et al. (1995) the uncertainties in the 
line fluxes are between 5 and 10 \%.}
\tablenotetext{b}{Internal extinction derived from the
H$\alpha$/H$\beta$ ratios for the optical nucleus, and from the near-IR
colors for the infrared nucleus.} \tablenotetext{c}{Observed H$\alpha$
flux, not corrected for extinction, in units of 10$^{-14}$ erg s$^{-1}$
cm$^{-2}$. The factor 3.2 discrepancy between the two values listed for
the optical nucleus are due to differences in the size of the region.
While this work gives the value for the central 1$'' \times
1''$, Kim's value corresponds to a region of 2$''$ wide along
the E$-$W direction.}

\tablenotetext{d}{The almost factor 2 discrepancy between the two
values for the optical nucleus is due to aperture effects (see above).
The larger value represents a more reliable value since it corresponds to
a smaller aperture centered on the optical peak  emission.}

\tablenotetext{e}{Extinction-corrected luminosity using the listed
A$_V$ values in units of erg s$^{-1}$ \AA$^{-1}$.}

\tablenotetext{f}{Emission line ratios. First row for our measurements
of the optical nucleus presents the observed ratios while the second
row gives the extinction-corrected values.  The values from Kim et al.
are extinction-corrected with the value given in the second column.
The ratios for the near-IR nucleus have not been corrected for internal
extinction.  In addition to the hydrogen Balmer lines, the
collisionaly-excited lines used are [\ion{O}{3}] 5007\AA, [\ion{O}{1}]
6300\AA, [\ion{N}{2}] 6584\AA, and the [\ion{S}{2}] doublet
6717,6731\AA.}

\enddata \end{deluxetable} \normalsize \newpage

\figcaption{ Spatial distribution of the spectra in the spectral range
6792 - 7045 \AA\ corresponding to the redshifted H$\alpha$+[NII]
emission line complex. Numbers indicate the fiber/spectrum position at
the slit/detector. These spectra are linearly scaled between their lowest and
highest values in the represented spectral range.}

\figcaption{{\it INTEGRAL} and {\it HST}/WFPC2 images in the central
regions of IRAS 17208-0014. Values for the isolines (in arbitrary
units) are: i) H$\alpha$ map:  2.024, 5.402, 11.04, 20.43, 36.11,
62.26, 105.9, 178.6, ii) Continuum map: 0.667, 1.333, 2.00, 2.667,
3.333, 4.00, 4.667, 5.334, 6.00 (contours in the upper-right,
lower-left, and lower-right correspond to the same intensity level),
iii) H$\alpha$/[NII] map: 0.443 (blue), 0.914, 1.386, 1.857, 2.328
(white). For the velocity maps, the scale is indicated at the right of
each panel. North is at the top and East is to the left.}

\figcaption{ NICMOS K image of IRAS 17208-0014 (first analysed by
Scoville et  al, 2000), superimposed on the ionized gas velocity field
(bottom) and the  velocity dispersion map (upper). The values for the
isolines are: i) velocity field:  12670, 12711, 12752, 12794, 12835,
12877, 12918, and 12960 km/s, ii) velocity  dispersion map:  240, 272,
304, 336, 368, 400 km/s.}

\newpage
\plotone{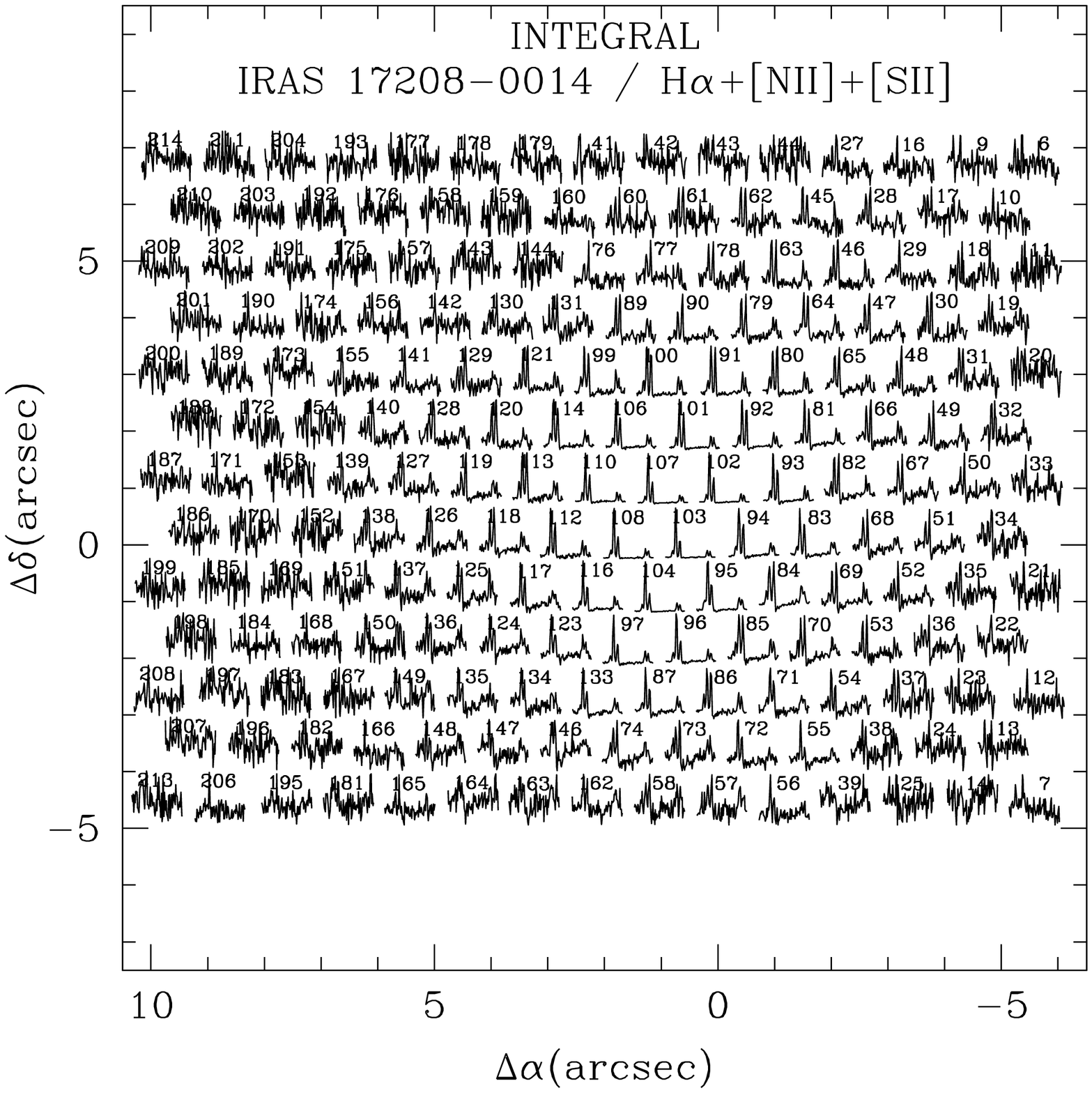}
\newpage
\plotone{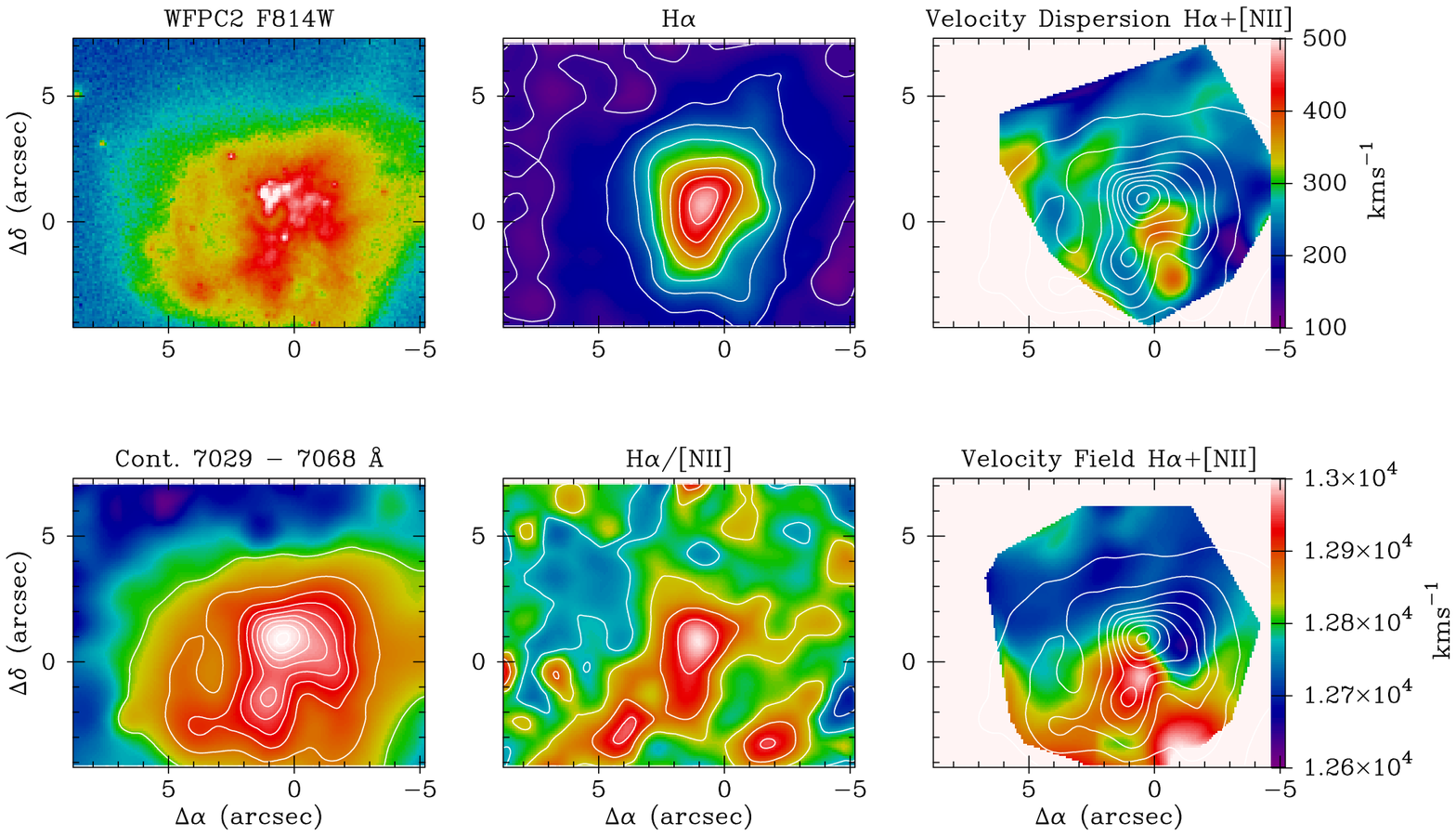}
\newpage
\plotone{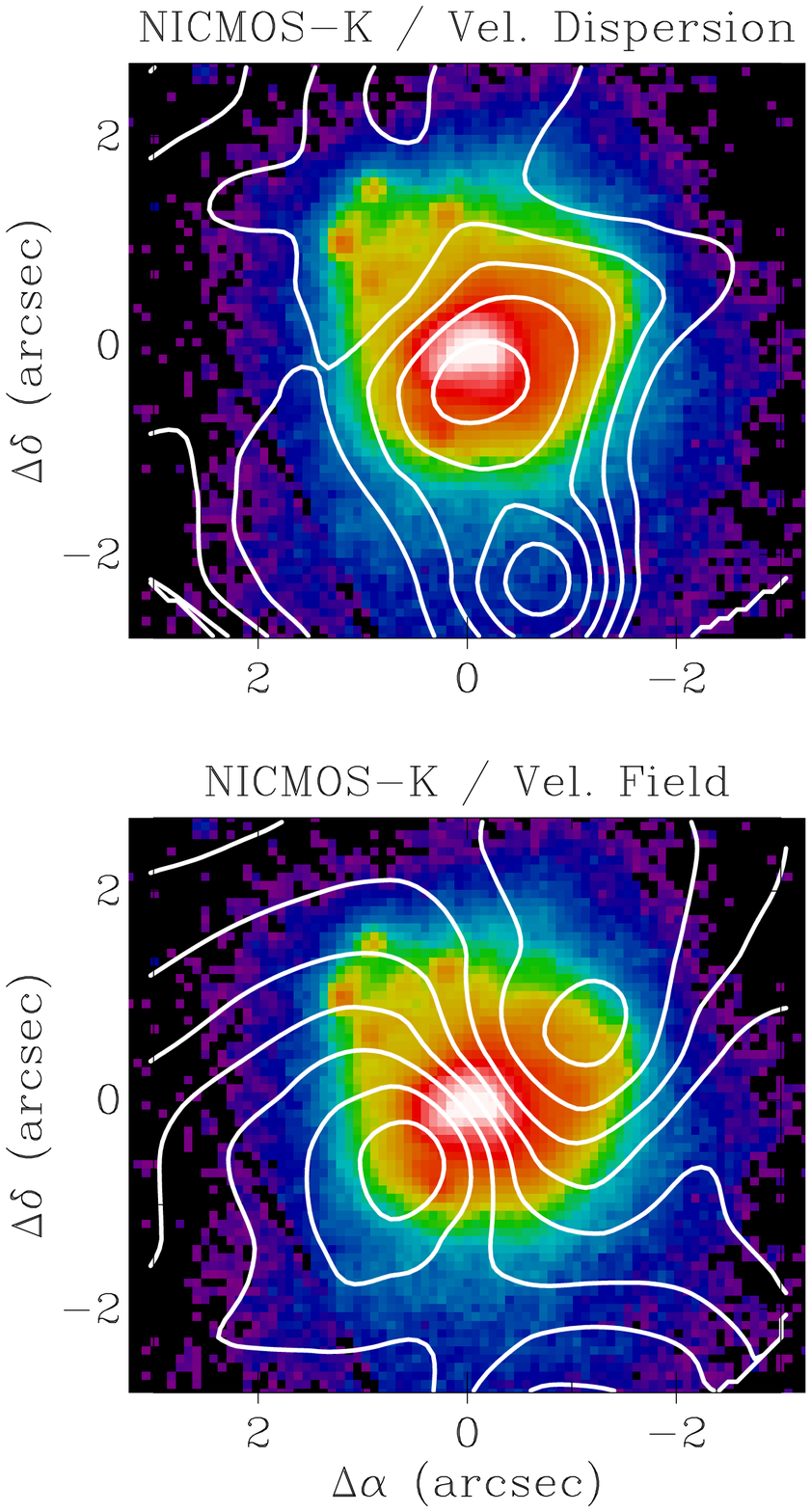}

\end{document}